\documentstyle[12pt]{article}

\newcommand{\beq}{\begin{equation}}
\newcommand{\eeq}{\end{equation}}
\newcommand{\beqa}{\begin{eqnarray}}
\newcommand{\eeqa}{\end{eqnarray}}
\newcommand{\beqar}{\begin{eqnarray*}}
\newcommand{\eeqar}{\end{eqnarray*}}

\newcommand{\Ga}{\Gamma}

\newcommand{\inn}{\!\cdot\!}

\newcommand{\z}{\zeta}

\newcommand{\eg}{{\it e.g.,}\ }
\newcommand{\ie}{{\it i.e.,}\ }
\newcommand{\labell}[1]{\label{#1}} %{\label{#1}} %
\newcommand{\reef}[1]{(\ref{#1})}
\newcommand\prt{\partial}

\begin{document}

\thispagestyle{empty} \rightline{\small hep-th/0304145 \hfill
IPM/P-2003/019} \vspace*{1cm}

\begin{center}
{\bf \Large Slowly varying tachyon and
 tachyon potential \\

 }
\vspace*{1cm}

{Mohammad R. Garousi}\\
\vspace*{1cm} { Department of Physics, Ferdowsi university, P.O.
Box 1436, Mashhad,
Iran}\\
\vspace*{0.1cm}  { Institute for Studies in Theoretical Physics
and Mathematics
IPM} \\
{P.O. Box 19395-5531, Tehran, Iran}\\
\vspace*{0.4cm}

\vspace{2cm} ABSTRACT
\end{center}
We show that   the scattering amplitude of  four open string
scalars or tachyons on the world-volume of a D$_p$-brane in the
bosonic string theory can be written in a universal form. The
difference between this amplitude and the corresponding amplitude
in the superstring theory is in an extra tachyonic pole. We show
that in an $\alpha'$ expansion and for slowly varying fields, the
amplitude is consistent with the tachyonic DBI action in which
the even part of the tachyon potential is  $V(T)=e^{-(\sqrt{\pi}
T/\alpha)^2}$ with $\alpha=1$ for bosonic theory and
$\alpha=\sqrt{2}$ for superstring theory. \vfill
\setcounter{page}{0} \setcounter{footnote}{0}
\newpage

\section{Introduction} \label{intro}
Decay of unstable branes is an interesting process which might
shed some new light in understanding properties of string theory
in time-dependent backgrounds \cite{mgas}-\cite{nlhl}. In
particular, it was pointed out by  Sen  that an effective action
of the Born-Infeld type proposed in \cite{mg1,ebmr,jk} can
capture many properties of these decaying processes \cite{asen2}.
This action for tachyon  and for the transverse scalar fields is
\cite{asen6,mg1,ebmr,jk}: \beqa S&=&-T_p\int d^{p+1}x
V(T)\sqrt{-\det(\eta_{ab}+2\pi\alpha' \prt_a\Phi^i\prt_b \Phi_i+
2\pi\alpha'\prt_a T\prt_b T)}\,\,, \label{dbiac} \eeqa where
$V(T)$ is the tachyon potential which is one at unstable vacuum
$T=0$ and should be zero at the true stable vacuum. Choosing for
the potential \cite{kkk,flaw,nlhl,dkvn,ko} \beqa
V(T)&=&1/\cosh(\sqrt{\pi}T/\alpha)\,\,,\labell{potential}\eeqa
 with
$\alpha=1$ for the bosonic theory and $\alpha=\sqrt{2}$ for the
superstring theory, one finds from above action the correct
stress-tensor in the homogenous time-dependent tachyon
condensation which is a tachyon solution that starts at the top
of the potential at $x^0\rightarrow -\infty$ and then rolls
toward  the true vacuum \cite{asen2,flanst,nlhl}. More over,  in
boundary conformal field theory, it has been shown  that the
tachyon action with the potential \reef{potential}  can be
reproduced by nearly on-shell S-matrix elements of slowly varying
tachyon vertex operators in the above time-dependent background
\cite{dkvn}.

It has been  shown in \cite{mg2} that all S-matrix elements
involving  four scalar and/or tachyon vertex operators in the
perturbative superstring theory can be written in a universal
form. The leading terms of the $\alpha'$ expansion of this
amplitude produces correctly the couplings of four slowly varying
scalar or tachyon fields as well as  a potential for tachyon
which is consistent with the following potential: \beqa V(T)&=&
e^{-(\sqrt{\pi}T/\alpha)^2}\,\,,\labell{potential1}\eeqa with
$\alpha=\sqrt{2}$.  In this paper, we extend this calculation to
the bosonic theory and show that the even part of the tachyon
potential in this case is again consistent with \reef{potential1}
with $\alpha=1$.

As a warm up in this introduction, we discuss the S-matrix element
of one closed string tachyon vertex operator and two open string
scalar or tachyon vertex operators. These amplitudes are
calculated in \cite{mggm}. They are\footnote{Using the fact that
the closed strings are not functional of the open string tachyon
but are functional of the scalar fields \cite{mg5}, we assumed
that the closed string field does not depend on the scalar. This
makes similar the S-matrix elements involving the scalars and the
tachyon as much as possible.}: \beqa A(\tau,
\Phi,\Phi)&\sim&\eta_{ij}\z_1^i\z_2^j\frac{\Gamma(-1-2s)}
{\Gamma(-s)\Gamma(-s)}\,\,,\nonumber\\
A(\tau,T,T)&=&
\frac{\Gamma(-1-2s)}{\Gamma(-s)\Gamma(-s)}\,\,,\nonumber\eeqa
where $\tau$ is the closed string tachyon. In the above equation
$s=-\alpha'(k_1+ k_2)^2/2$, $\z_1^i,\z_2^i$ are the scalar
polarization vectors in the $(26-(p+1))$-dimensional transverse
space, and  $k_1^a,k_2^a$ are the  momentum vectors in the
$(p+1)$-dimensional world-volume space\footnote{Our index
conventions are that early Latin indices take values in the
world-volume, \eg $a,b=0,1,...,p$, middle Latin indices take
values in the transverse space, \eg $i,j=p+1,...,24,25$, early
Greek indices take values in the (27-(p+1))-dimensional space, \eg
$\alpha,\beta=p+1,...,24,25,26$, and middle Greek indices take
value in the whole 26-dimensional space of bosonic theory, \eg
$\mu,\nu=0,1,...,24,25$ }. The momenta in this amplitude satisfy
the on-shell condition $k^2=0$ for the scalar and $k^2=1/\alpha'$
for the tachyon. Now as can be seen, the two amplitudes has the
same dependency on the Mandelstam variable $s$. To write both in a
universal form, we introduce the polarization vectors
$\z_1^{\alpha},\z_2^{\beta}$ in a $(27-(p+1))$-dimensional space
with $\z^{26}$ polarization of the tachyon.  In terms of these new
polarization vectors, the universal amplitude is, \beqa
A(\z_1,\z_2)&\sim&\z_1\inn
\z_2\left(\frac{\Ga(-1-2s)}{\Ga(-s)\Ga(-s)}\right)
\,\,.\labell{axxon}\eeqa  The metric used to multiply the
polarizations is $\eta^{\alpha\beta}$.  Note that if we do not
use the on-shell condition for the momenta, the amplitude in this
form is invariant under $SO(p+1)\times SO(27-(p+1))$. Comparing
this amplitude with the corresponding amplitude in the
superstring theory \cite{mg2}, one finds the following relation:
\beqa A^{{\rm bosonic \,string }}&=&\frac{-1}{1+2s}A^{{\rm
suprstring}} \,\,.\eeqa

We are interested in the $\alpha'\rightarrow 0$ limit of S-matrix
elements.  To do this expansion, one should first off-shell extend
the amplitude, \ie not using on-shell conditions for the momenta,
then  expand  the gamma functions. At the end one  restricts the
momenta to the  on-shell conditions. The on-shell amplitudes have
in general two kinds of terms upon expanding the amplitude at
this limit. One includes terms like $k_i\inn k_j$ that do not
depend on the on-shell condition, and the other kind includes
term like $k_i^2$ that highly depend on the on-shell condition.
Only these latter terms are different for scalars and for
tachyons. So one may separate the amplitude after $\alpha'$
expansion into  two parts, \ie $ A=A'+A^{{\rm on}}$. The first
part is the same for scalar and tachyon, whereas, the second part
depends on open string states, \eg $A^{{\rm on}}=0$ when all
states are scalars. Our speculation \cite{mg1,mg3,mg2} is that
while the leading $\alpha'$ order terms of $A$ for only scalars
are correspond to DBI action,  the same $A$ for scalars and/or
tachyons are correspond to tachyonic DBI action \reef{dbiac} with
potential \reef{potential1}.

To $\alpha'$ expand the amplitude \reef{axxon},  one should first
extend it to off-shell physics. One may write the off-shell
amplitude as the following: \beqa A^{{\rm off}}&\sim&\z_1\inn
\z_2\left(\left(\frac{\Ga(-1-2s)}{\Ga(-s)\Ga(-s)}-G\right)+G
\right) \,\,,\labell{axxoff}\eeqa where the momenta in this
action do not  satisfy the on-shell condition. The function
$G(k_1,k_2)$ has to cancel non-desirable terms producing by
expansion of the gamma functions in the $\alpha'\rightarrow 0$
limit, \ie the terms in the first parenthesis being independent of
 $k_1^2$, or $k_2^2$. Since we are interested in slowly varying
fields we continue our discussion for the leading term of
expansion. For this term the function $G$ should be \ie
 \beqa G&=&
-\alpha'(k_1^2+k_2^2)/4+\cdots\,\,.\nonumber\eeqa   Now $\alpha'$
expand the gamma functions in the amplitude, and then go back to
the on-shell physics, \ie  \beqa A&\sim&\z_1\inn
\z_2\left(\frac{}{}\left(-\alpha'k_1\inn
k_2/2+\cdots\right)+G^{{\rm on}} \right)\,\,,\labell{expand}\eeqa
where dots represent non-leading terms. The function $G^{{\rm
on}}$ is $G$ in which the momenta are on-shell, \ie $G^{{\rm
on}}=0$ when the two states are scalars and $G^{{\rm on}}=-1/2$
when the two states ate tachyons. Now compare it with expansion
of the tachyonic DBI action \reef{dbiac}\footnote{We have used
the fact that the closed string tachyon  normalizes the D-brane
tension as $T_p\rightarrow T_p(1+\tau+\cdots)$\cite{mggm}, and we
have kept only linear term for the closed string tachyon.}. The
action \reef{dbiac} has the following expansion: \beqa
S&=&-T_p\int d^{p+1}x\,\tau\left(1-\pi T^2+\pi \alpha'(\prt
 T)^2+\pi \alpha'(\prt \Phi)^2\cdots\right)\,\,,\nonumber\eeqa
 where  we have used the
expansion $V(T)=1-\pi T^2+O(T^4)$ for the even part of tachyon
potential. When the polarization vectors in \reef{expand} are in
the transverse directions, \ie in the $(26-(p+1))$-dimensional
space, the first term of \reef{expand}
 reproduces the  kinetic term of the scalar fields. When the polarization
vectors are in the 27-th direction, the first term is reproduced
by the kinetic term of tachyon,  and  $G^{{\rm on}}$ term is
reproduced by the tachyon potential. In the present paper we
would like to extend these calculation to the case of S-matrix
elements of four tachyon and/or scalar vertex operators.

In next section , we describe the calculation of the scattering
amplitudes in the bosonic string theory, and write the results in
a universal amplitude which is invariant under $SO(p+1)\times
SO(27-(p+1))$ when the momenta are off-shell. Then in the section
3, using the idea in \cite{mg2}, we
 off-shell extend the amplitude,  expand it in the limit
$\alpha'\rightarrow 0$, and then back to the on-shell amplitude.
Then compare the leading terms of the expansion with the action
\reef{dbiac} and find the coefficient of  $T^4$ in the expansion
of tachyon potential in this action.

\section{Scattering amplitude} Scattering amplitude of four
scalar or tachyon  vertex operators in the bosonic string  theory
is given by the following correlation function:
 \beqa A&\sim&\int dx_1dx_2dx_3dx_4\nonumber\\
&&\times\langle:V(2\inn k_1,x_1):V(2\inn k_2,x_2):V(2\inn
k_3,x_3):V(2\inn k_4,x_4):\rangle \,\,,\labell{amp}\eeqa where
\beqa V^{{\rm
tachyon}}&=&\z^{26} e^{2k\cdot  X}\,\,,\nonumber\\
V^{{\rm scalar}}&=&\z^i\prt X_i e^{2k\cdot  X} \,\,,\nonumber\eeqa
with the world-sheet propagator \beqa \langle
X^{\mu}(x)X^{\nu}(y)\rangle&=&-\alpha'\eta^{\mu\nu}\ln(x-y)/2
\,\,,\nonumber \eeqa where we have used the doubling trick to
work with only holomorphic  functions on the boundary of
world-sheet \cite{mgrm}. Using the Wick theorem, one can easily
calculate the correlators in \reef{amp}. The different between
the tachyon and  scalar vertex operators is in the extra factor
of $\prt X^i$. However, because the momenta in the vertex
operators are in the world-volume directions and index $i$ takes
value in the transverse space, the factors $\prt X^i$ correlate
only among themselves resulting contraction of the scalar
polarizations. Performing the correlation, one finds that the
integrand has $SL(2,R)$ symmetry. One should fix this symmetry by
fixing position of three vertices in the real line. Different
fixing of these positions give different ordering of the four
vertices in the boundary of the world-sheet. One should add all
non-cyclic permutation of the vertices to get the correct
scattering amplitude. So one should add the amplitudes resulting
from the fixing $(x_1=0, x_2, x_3=1,x_4=\infty)$, $(x_1=0,
x_2,x_4=1,x_3=\infty)$, $(x_1=0, x_3,x_4=1,x_2=\infty)$,
$(x_1=0,x_3,x_2=1,x_4=\infty)$, $(x_1=0, x_4, x_2=1,x_3=\infty)$,
$(x_1=0,x_4,x_3=1,x_2=\infty)$. After these gauge fixing, one
ends up with only one integral  which gives the beta function. The
resulting amplitude for scalars and tachyons seems different,
however, using the conservation  of momenta, one can write them
all in a universal form:
 \beqa
A(\z_1,\z_2,\z_3,\z_4)&=&\,A_s(\z_1,\z_2,\z_3,\z_4)+\,A_u
(\z_1,\z_2,\z_3,\z_4)+ \,A_t(\z_1,\z_2,\z_3,\z_4)\,\,,
\nonumber\eeqa where $A_s,A_u$, and $A_t$ are the part of the
amplitude that  has  infinite tower of poles in $s$-, $u$- and
$t$-channels, respectively. They are
 \beqa A_s&= &4iT_p\z_1\inn\z_2\,\z_3\inn\z_4
\left(\frac{}{}\frac{\Ga(-1-2s)\Ga(1+s+u-t)}{\Ga(u-t-s)}
\right.\nonumber\\
&&\left.+\frac{\Ga(-1-2s)\Ga(1+t+s-u)}{\Ga(t-u-s)}+
\frac{\Ga(1+s+t-u)\Ga(1+s+u-t)}{\Ga(2+2s)} \frac{}{} \right)\,\,,
\nonumber\\
A_u&=&4iT_p\z_1\inn\z_3\,\z_2\inn\z_4\left(\frac{}{}
\frac{\Ga(1+t+u-s)\Ga(1+s+u-t)}{\Ga(2+2u)}\right.\nonumber\\
&&\left.+\frac{\Ga(-1-2u)\Ga(1+t+u-s)}{\Ga(t-s-u)}
+\frac{\Ga(-1-2u)\Ga(1+s+u-t)}{\Ga(s-t-u)} \frac{}{}
\right)\,\,,\nonumber\\
A_t&=&4iT_p\z_1\inn\z_4\,\z_2\inn\z_3\left(\frac{}{}
\frac{\Ga(-1-2t)\Ga(1+t+u-s)}{\Ga(u-t-s)}\right.\nonumber\\
&&\left.+ \frac{\Ga(1+t+u-s)\Ga(1+t+s-u)}{\Ga(2+2t)}
+\frac{\Ga(-1-2t)\Ga(1+t+s-u)}{\Ga(s-t-u)} \frac{}{}
\right)\,\,,\labell{aton}\eeqa where $\z_i$'s are the
polarization vectors of the scalars-tachyon, and the Mandelstam
variables $s,t,u$ are the following:\beqa
s&=&-\alpha'(k_1+k_2)^2/2\,\,,\nonumber\\
t&=&-\alpha'(k_2+k_3)^2/2\,\,,\labell{mandel}
\\
u&=&-\alpha'(k_1+k_3)^2/2\,\,.\nonumber \eeqa Using conservation
of momenta, one finds that they satisfy the relation \beqa
s+t+u&=&-\alpha'(\sum_{i=1}^{4}k_i^2)/2\,\,,\labell{mandel1}\eeqa
where $k^2=0$ for on-shell scalar and $k^2=1/\alpha'$ for the
on-shell tachyon.
 We have also
normalized the amplitude \reef{aton} by the factor $2iT_p$ to
have the complete agreement with the tachyonic DBI action
including the normalization of fields. Note that if one dose not
use on-shell conditions for the momenta, the amplitude \reef{aton}
will be invariant under $SO(p+1)\times SO(27-(p+1))$. Also the
amplitude $A_s$ is symmetric under $1\leftrightarrow
2,3\leftrightarrow 4$, the amplitude $A_u$ is symmetric under
$1\leftrightarrow 3,2\leftrightarrow 4$, the amplitude $A_t$ is
symmetric under $1\leftrightarrow 4,2\leftrightarrow 3$, and the
whole amplitude $A$ is symmetric under interchanging any particle
label. Comparing this amplitude with the corresponding amplitude
in the supestring theory \cite{mg2}, one observes the following
relation \beqa A_s^{{\rm
bosonic\,string}}&=&\frac{-1}{1+2s}A_s^{{\rm supestring}}\,\,,\nonumber\\
A_u^{{\rm
bosonic\,string}}&=&\frac{-1}{1+2u}A_u^{{\rm supestring}}\,\,,\nonumber\\
A_t^{{\rm bosonic\,string}}&=&\frac{-1}{1+2t}A_t^{{\rm
supestring}}\,\,.\nonumber\eeqa Extension of the scattering
amplitude \reef{aton} to the noncommutative case is very easy.
One should add just some phase factors to the amplitude, and use
the open string metric in \reef{mandel} \cite{ewns}. The above
relation between the amplitude in bosonic and super string theory
remains the same. So one may use the noncommutative results in the
superstring theory \cite{mg2} and just add the extra tachyonic
pole to find the result in the bosonic theory.

\section{The $\alpha'$ expansion and effective action}
Now we would like to expand this amplitude at the limit
$\alpha'\rightarrow 0$. To do this, we should  analytically
continue the amplitude to the off-shell physics, \eg  the momenta
do not satisfy the on-shell condition, then expand the gamma
functions. After finishing with the $\alpha'$ expansion we go
back to the on-shell physics. In this way, one finds massless
pole and infinite number of contact terms. It was shown in
\cite{mg2} that for the non-BPS brane of supestring theory, the
massless pole and contact terms of noncommutative tachyonic DBI
action are produced by  the universal string theory amplitude
expanded  at $\alpha'\rightarrow 0$.

The amplitudes $A_s$, $A_u$ can be read from the amplitude $A_t$
by changing the states label, \ie replacing $(1432)\rightarrow
(1324)$  one gets $A_u$, and replacing $(1423)\rightarrow (1234)$
one gets $A_s$.  So we continue our discussion only for the $A_t$
amplitude. Off-shell extension of $A_t$ is  \beqa A_t^{\rm
off}\!\!\! &=&\!\!\!4iT_p\z_1\inn\z_4\,\z_2\inn\z_3 \left(\left(
\frac{\Ga(-1-2t)\Ga(1+t+u-s)}{\Ga(u-s-t)}\right.\right.\labell{atoff}\\
\!\!\!\!\!&&\!\!\!\!\!+\left.\left.\frac{\Ga(1+t+u-s)\Ga(1+s+t-u)}{\Ga
(2+2t)}
+\frac{\Ga(-1-2t)\Ga(1+t+s-u)}{\Ga(s-u-t)}-3F_t\right)+3F_t
\frac{}{} \right),\nonumber\eeqa where the function
$F_t(k_1,k_2,k_3,k_4)$ has  to  cancel the non-desirable
$\alpha'^2$ order  terms  resulting from expansion of the gamma
functions in the low energy limit, \ie the first parenthesis must
be independent of $k_i^2$ for any $i=1,2,3,4$.

Aside  from the tachyonic pole in the gamma functions in
\reef{atoff}, these functions have the following expansion   at
$\alpha'\rightarrow 0$:
 \beqa
\frac{\Ga(-1-2t)\Ga(1+t+u-s)}{\Ga(u-s-t)}&=&\frac{-1}{1+2t}\left(
\frac{1}{2}+\frac{s-u}{2t}-
\frac{\pi^2}{6}\left((s-u)^2-t^2\right)+\cdots\right)\,\,,\nonumber\\
\frac{\Ga(-1-2t)\Ga(1+t+s-u)}{\Ga(s-u-t)}&=&\frac{-1}{1+2t}\left(
\frac{1}{2}+\frac{u-s}{2t}-
\frac{\pi^2}{6}\left((s-u)^2-t^2\right)+\cdots\right)\,\,,\nonumber\\
\frac{\Ga(1+t+u-s)\Ga(1+s+t-u)}{\Ga(2+2t)}&=&
\frac{-1}{1+2t}\left(-1-\frac{\pi^2}{6}\left((s-u)^2-t^2\right)+
\cdots\right)\,\,,\labell{gamma}\eeqa  where dots represent terms
that have the Madelstam variables at lest cubic.  Expansion of
the tachyonic pole is also $1/(1+2t)=1-2t+O(t^2)$. Unlike the
noncommutative case that each gamma function carries a phase
factor \cite{mg3,mg2}, the gamma functions in the commutative case
\reef{atoff} have no such  factors. Hence, the constant and the
massless pole terms in \reef{gamma} will be canceled when
inserted into \reef{atoff}. Up to terms of second order of the
Mandelstam variables, one then left with exactly the same terms
that appears in the superstring case. So  function $F_t$ is
exactly the one appears in the superstring theory \cite{mg3},\beqa
F_t&=&-\frac{\alpha'\pi^2}{6}
\left(-\frac{\alpha'}{4}(\sum_{i=1}^4
k_i^2)^2-t\sum_{i=1}^4k_i^2-\frac{\alpha'}{2}(k_1^2+k_4^2)(k_2^2+k_3^2)
\right.\nonumber\\
&&\left.
+\frac{\alpha'}{2}(k_1^2+k_2^2)(k_3^2+k_4^2)+\frac{\alpha'}{2}
(k_1^2+k_3^2)(k_2^2+k_4^2) \right)\,\,.\labell{F}\eeqa   We have
only rearranged terms  to have the symmetry $1\leftrightarrow 4$,
$2\leftrightarrow 3$ manifest. Now at this point  one should
restrict the momenta to the on-shell to return to the on-shell
amplitude. The resulting amplitude is
 \beqa A_t&\!\!\!=\!\!\!&
-12iT_p\,\z_1\inn\z_4\,\z_2\inn\z_3\labell{ac41}\\
&&\times\left({\pi^2\alpha'^2\over 3}(k_2 \inn k_3)(k_1\inn k_4)
-\frac{\pi^2\alpha'^2}{3}(k_1\inn k_2)(k_3\inn
k_4)-\frac{\pi^2\alpha'^2}{3}(k_1\inn k_3)(k_2\inn k_4)-F_t^{{\rm
on}}\frac{}{}\right),\nonumber\eeqa
 plus some other terms
that are at least cubic in the Mandelstam variables.  These terms
are related to the higher derivative terms in the field theory.
Note that for the four scalar case, the higher derivative terms
are also of higher power of $\alpha'$ relative to the terms in
\reef{ac41}. However, when one consider four tachyon amplitude,
these higher derivative terms might be of the same order of
$\alpha'$ as those  appearing in \reef{ac41}, \eg $(\prt_a
T\prt^aT)^2$ and $T^2\prt_a\prt_bT\prt^a\prt^bT$ are both of the
same order of $\alpha'$. However, the second one is very small
relative to the first one for slowly varying tachyon field, in
which we are nor interested in this paper. In the above couplings
$F_t^{{\rm on}}$ means $F$ in which the momenta are on-shell, \ie
\beqa F_t^{{\rm on}}=0\quad; &{\rm
for}&\Phi_1\Phi_2\Phi_3\Phi_4\,\,,\nonumber\\
F_t^{{\rm on}}=-\frac{\pi^2}{3}\alpha'k_2\inn k_3\quad;&{\rm
for}&T_1\Phi_2\Phi_3T_4\,\,,\nonumber\\
F_t^{{\rm on}}=-\frac{\pi^2}{3}(\alpha'k_2\inn k_3+\alpha' k_1\inn
k_4+1)\quad;&{\rm for }&T_1T_2T_3T_4\,\,.\labell{ft}\eeqa

As mentioned before, apart from the on-shell function $F_t^{{\rm
on}}$, the amplitude \reef{ac41} is the same for scalars and the
tachyons. So the field theory consistent with this part must
contain both scalar and the tachyon in the same footing. However,
 $F_t^{{\rm on}}$ is different for the scalars and for the
tachyons. So this term is responsible for the extra terms that the
field theory produces  due to  the tachyon potential. Now it is
easy to replace the values of $F_t^{{\rm on}}$ into \reef{ac41},
add the contribution from $A_s$, $A_u$, \ie $A=A_t+A_s+A_u$, and
compare the whole amplitude with the tachyonic action
\reef{dbiac}. For the case that the four states are tachyons, the
amplitude  simplifies to: \beqa A\!\!\!&=&\!\!\!
-16\pi^2iT_p\z_1^{26}\z_2^{26}\z_3^{26}\z_4^{26}\left(-\frac{\alpha'^2}{4}
(k_2\inn k_3)(k_1\inn k_4) -\frac{\alpha'^2}{4}(k_1\inn
k_2)(k_3\inn
k_4)-\frac{\alpha'^2}{4}(k_1\inn k_3)(k_2\inn k_4)\right.\nonumber\\
\!\!\!\!\!&&\!\!\!\!\!\left.+ \frac{\alpha'}{4}(k_2\inn
k_3+k_1\inn k_4+k_1\inn k_3+k_2\inn k_4+k_1\inn k_2+k_3\inn k_4)+
\frac{3}{4}\right)\,.\labell{ac43}\eeqa Compare it with the
following expansion of action \reef{dbiac}: \beqa S &=&-T_p\int
d^{p+1}x
\left(1-\pi T^2 +\beta T^4\right.\nonumber\\
&&\left.+\pi \alpha'(\prt_a T\prt^a T)
-\pi^2\alpha'T^2(\prt_aT\prt^aT)-
\frac{\pi^2\alpha'^2}{2}(\prt_aT\prt^aT)^2+\cdots\right)\nonumber\,\,,
\labell{ltttt}\eeqa where  the constant  $\beta$ is the
coefficient of $T^4$ in the tachyon potential. One can easily
observe that the term with four derivatives reproduces exactly
the four momentum contact terms in \reef{ac43}, term with two
derivatives reproduces the two momentum contact terms, and the
$T^4$ term produces the constant term in \reef{ac43} provided
that $\beta=\pi^2/2$.  So the even part of tachyon potential
expanded around its maximum, \ie around $T_{max}=0$, has  the
following expansion: \beqa V(T)=1-\pi
T^2+\frac{\pi^2}{2}T^4+O(T^6)\,\,.\labell{finalv} \eeqa This
expansion is consistent with the potential \reef{potential1} with
$\alpha=1$.

Consider the following observations: 1) the off-shell S-matrix
elements involving the transverse scalars and tachyon  are
invariant under $SO(p+1)\times SO(d+1-(p+1))$ where $d$ is
critical dimension of theory,   2) the  kinetic term of both
scalar and tachyon appears in the same footing in the effective
action \reef{dbiac}. Using these,  one may suspect that the
tachyon, like the scalar field, represents a new physical
dimension in the low energy physics. The idea that tachyon might
represent a new dimension was  remarked also in
\cite{ph,ebmr}\footnote{See \cite{khsn} for a possible
geometrical meaning of the tachyon in the intersecting brane
antibrane.}. Using this idea one may try to write a covariant
action as:\beqa S&=&-T_p\int d^{p+1}\sigma
V(X^2)\sqrt{-\det(P[\eta_{ab}])}\,\,,\eeqa where $X^A$ is the
(d+1)-dimensional contra-variant vector, and the pull-back is
\beqa P[\eta_{ab}]&=&\eta_{AB}\frac{\prt X^A}{\prt
\sigma^a}\frac{\prt X^B}{\prt \sigma^b}\,\,.\nonumber \eeqa This
action has global $SO(d+1)$ symmetry, and the gauge symmetry of
world-volume coordinate transformation. Using this latter
symmetry, one can chose the static gauge, \ie $X^a=\sigma^a$.
Letting $X^i=2\pi\alpha' \Phi^i\,,X^{d+1}=2\pi\alpha' T$, the
pull-back becomes \beqa P[\eta_{ab}]&=&
\eta_{ab}+2\pi\alpha'\prt_a\Phi^i\prt_b\Phi_i+2\pi\alpha'\prt_aT\prt_b
T\,\,.\nonumber\eeqa The action in this gauge has global
$SO(p+1)\times SO(d+1-(p+1))$ symmetry, however, because the
actual on-shell potential depends only on the (d+1)-th
coordinate, \ie $V(X^2)\rightarrow V(T^2)$, this symmetry is
broken to $SO(p+1)\times SO(d-(p+1))$ in the on-shell physics.

{\bf Acknowledgement}: I would like to thank M. Alishahiha  for
discussion.

%\newpage


\begin{thebibliography}{99}

\bibitem%[mgas]
{mgas}{M. Gutperle and A. Strominger, `` Spacelike branes,'' IHEP
{\bf 0204},018 (2002) [hep-th/0202210].}
\bibitem%[asen1]
{asen1}{A. Sen, ``Rolling tachyon,'' JHEP {\bf 0204}, 048 (2002)
[hep-th/0203211].}
\bibitem%[asen2]
{asen2}{A. Sen,  ``Tachyon Matter,'' JHEP {\bf 0207}, 065 (2002)
[hep-th/0203265].}
\bibitem%[asen3]
{asen3}{A. Sen, ``Field theory of tachyon matter,'' Mod. Phys.
Lett. A {\bf 17}, 1797 (2002) [hep-th/0204143].}
\bibitem%[asen4]
{asen4}{A. Sen, ``Time evolution in open string theory,'' JHEP
{\bf 0210}, 003 (2002) [hep-th/0207105].}
\bibitem%[asen5]
{asen5}{A. Sen, ``Time and tachyon,'' [hep-th/02091220.}
\bibitem%[pmas]
{pmas}{P. Mukhopadhyay and A. sen, ``Decay of unstable D-brane
with electric field, '' JHEP {\bf 0211}, 047
(2002)[hep-th/0208142].}
\bibitem%[as]
{as}{A. Strominger, ``Open string creation by S-branes,''
[hep-th/0209090].}
\bibitem%[flanst]
{flanst}{F. Larsen, A. Naqvi and S. Terashima, ``Rolling tachyons
and decaying branes, '' JHEP {\bf 0302}, 039 (2003)
[hep-th/0212248].}
\bibitem%[mgas1]
{mgas1}{M. Gutperle and A. Strominger, ``timelike Boundary
Liouville Theory,'' [hep-th/0301038].}
\bibitem%[amas]
{amas}{A. Maloney, A. Strominger and X. Yin, ``S-brane
thermodynamica,'' [hep-th/0302146].}
\bibitem%[toss]
{toss}{T. Okuda and S. Sugimoto, ``Coupling of rolling tachyon to
closed strings,'' Nucl. Phys. B {\bf 647}, 101 (2002)
[hep-th/0208196].}
\bibitem%[srss]
{srss}{S. J. Rey and S. Sugimoto, ``Rolling tachyon with electric
and magnetic fields: T-duality approach,'' [hep-th/0301049].}
\bibitem%[srss1]
{srss1}{S. J. Rey and S. Sugimoto, ``Rolling of modulated tachyon
with gauge flux and emergent fundamental string,''
[hep-th/0303133].}
\bibitem%[nlhl]
{nlhl}{N. Lambert, H. Liu and J. Maldacena, ``Closed strings from
decaying D-branes,'' [hep-th/0303139].}
\bibitem%[mg1]
{mg1}{M. R. Garousi, ``Tachyon couplings on non-BPS D-branes and
Dirac-Born-Infeld action,''  Nucl. Phys. B {\bf 584}, 284 (2000)
[hep-th/0003122].}
\bibitem%[ebmr]
{ebmr}{E. A. Bergshoeff, M. de Roo, T. C. de Wit, E. Eyras and S.
Panda, ``T-duality and action for non-BPS D-branes,'' JHEP {\bf
0005}, 009 (2000) [hep-th/0003221].}
\bibitem%[jk]
{jk}{ J. Kluson, ``Proposal for non-BPS D-brane action,'' Phys.
Rev. D {\bf 62}, 126003 (2000) [hep-th/0004106].}
\bibitem%[asen6]
{asen6}{A. Sen, ``Supersymmetric world-volume action for non-BPS
D-branes,'' JHEP {\bf 9910}, 008 (1999) [hep-th/9909062].}
\bibitem%[kkk]
{kkk}{C.-J Kim, H.B. Kim, Y.-B Kim and O.-K Kim, ``Electromagnetic
string fluid in rolling tachyon,'' JHEP {\bf 0303} (2003) 008
[hep-th/0301076]; ``Cosmology of rolling tachyon,''
hep-th/0301142.}
\bibitem%[flaw]
{flaw}{F. Leblond and A. Peet, ``SD-brane gravity fields and
rolling tachyona,'' [hep-th/0303035].}

\bibitem%[dkvn]
{dkvn}{D. Kutasov and V. Niarchos, ``Tachyon Effective Actions In
Open String Theory,'' [hep-th/0304045].}
\bibitem%[ko]
{ko}{K. Okuyama, ``Wess-Zumino Term in Tachyon Effective
Action,'' [hep-th/0304108].}
\bibitem%[mg2]
{mg2}{M. R. Garousi, ``Off-shell extension of S-matrix elements
and tachyonic effective actions,'' [hep-th/0303239].}
\bibitem%[mggm]
{mggm}{M. R. Garousi and G. R. Maktabdaran, ``Excited D-brane
decay in cubic string field theory and in bosonic string
theory,'' Nucl. Phys. B {\bf 651}, 26 (2003) [hep-th/0210139].}
\bibitem%[mg3]
{mg3}{M. R. Garousi, ``On-shell S-matrix and tachyonic effective
actions,'' Nucl. Phys. B {\bf 647}, 117 (2002) [hep-th/0209068].}
\bibitem%[mgrm]
{mgrm}{M. R. Garousi and R. C. Myers, ``Superstring scattering
from D-branes,''  Nucl. Phys. B {\bf 475}, 193 (1996)
[hep-th/9603194].}
\bibitem%[ewns]
{ewns}{N. Seiberg and E. Witten, ``String Theory on
Noncommutative Geometry,'' JHEP {\bf 9909}, 32 (1999)
[hep-th/9908142].}
\bibitem%[ph]
{ph}{P. Horava, ``Type IIA D-Branes, K-theory and Matrix
Theory,'' Adv. Theor. Math. Phys. {\bf 2}, 1373 (1998)
[hep-th/9812135].}
\bibitem%[mg5]
{mg5}{M. R. Garousi and R. C. Myers, ``World-Volume Interactions
on D-Branes,''  Nucl. Phys. B {\bf 542} (1999) 73
[raXiv:hep-th/9809100].}
\bibitem%[khsn]
{khsn}{K. Hashimoto and S. Nagaoka, ``Recombination of
intersecting D-brane by local tachyon condensation,''
hep-th/0303204.}

\end{thebibliography}
\end{document}